\documentclass[prl,twocolumn,showpacs,superscriptaddress]{revtex4}
\usepackage{graphicx}
\usepackage{amsmath}
\usepackage{bm}
\usepackage[T1]{fontenc}
\date{\today}
\begin{document}
\title{Anomalous Hall effect in field-effect structures of (Ga,Mn)As}
\author{D. Chiba}
\affiliation{Semiconductor Spintronics Project, Exploratory Research for Advanced Technology, Japan Science and Technology Agency, Sanban-cho 5, Chiyoda-ku, Tokyo 102-0075, Japan}
\affiliation{Laboratory for Nanoelectronics and Spintronics$,$ Research Institute of Electrical Communication, Tohoku University, Katahira 2-1-1, Aoba-ku, Sendai, Miyagi 980-8577, Japan}
\affiliation{Institute for Chemical Research, Kyoto University, Gokasho, Uji, Kyoto 611-0011, Japan}
\author{A. Werpachowska}
\affiliation{Institute of Physics, Polish Academy of Sciences, PL-02 668 Warszawa, Poland}
\author{M. Endo}
\affiliation{Laboratory for Nanoelectronics and Spintronics$,$ Research Institute of Electrical Communication, Tohoku University, Katahira 2-1-1, Aoba-ku, Sendai, Miyagi 980-8577, Japan}
\author{Y. Nishitani}
\affiliation{Laboratory for Nanoelectronics and Spintronics$,$ Research Institute of Electrical Communication, Tohoku University, Katahira 2-1-1, Aoba-ku, Sendai, Miyagi 980-8577, Japan}\author{F. Matsukura}
\affiliation{Laboratory for Nanoelectronics and Spintronics$,$ Research Institute of Electrical Communication, Tohoku University, Katahira 2-1-1, Aoba-ku, Sendai, Miyagi 980-8577, Japan}
\affiliation{Semiconductor Spintronics Project, Exploratory Research for Advanced Technology, Japan Science and Technology Agency, Sanban-cho 5, Chiyoda-ku, Tokyo 102-0075, Japan}
\author{T. Dietl}
\affiliation{Institute of Physics, Polish Academy of Sciences, PL-02 668 Warszawa, Poland}
\affiliation{Institute of Theoretical Physics, University of Warsaw, ul. Ho\.za 69, PL-00 681 Warszawa, Poland}
\author{H. Ohno}
\affiliation{Laboratory for Nanoelectronics and Spintronics$,$ Research Institute of Electrical Communication, Tohoku University, Katahira 2-1-1, Aoba-ku, Sendai, Miyagi 980-8577, Japan}
\affiliation{Semiconductor Spintronics Project, Exploratory Research for Advanced Technology, Japan Science and Technology Agency, Sanban-cho 5, Chiyoda-ku, Tokyo 102-0075, Japan}
\email{ohno@riec.tohoku.ac.jp}

\begin{abstract}
The anomalous Hall effect in metal-insulator-semiconductor structures having thin (Ga,Mn)As layers as a channel has been studied in a wide range of Mn and hole densities changed by the gate electric field. Strong and  unanticipated temperature dependence, including a change of sign, of the anomalous Hall conductance $\sigma_{xy}$ has been found in samples with the highest Curie temperatures.  For more disordered channels, the scaling relation between $\sigma_{xy}$ and $\sigma_{xx}$, similar to the one observed previously for thicker samples, is recovered.
\end{abstract}
\maketitle

Along with anisotropic magnetoresistance, the anomalous Hall effect (AHE) results from an interplay between spin-orbit interactions and spin polarization of electric current specific to ferromagnets. It has been recently realized that for a certain region of conductivities, the anomalous Hall conductivity $\sigma_{xy}$ is a measure of the Berry phase of carrier trajectories in the $k$ space and, thus, provides information on the hitherto inaccessible aspects of the band structure topology in the presence of various spin-orbit interactions \cite{Sundaram:1999_a,Jungwirth:2002_a,Jungwirth:2006_a,Nagaosa:2009_a,Werpachowska:2009_a}. Interestingly, the effect appears to be qualitatively immune to disorder,  except for the case of linear-in-$k$ Rashba-type Hamiltonians in two-dimensional electron systems, where the contribution to $\sigma_{xy}$ vanishes \cite{Rashba:2004_d} unless the lifetime is spin-dependent \cite{Inoue:2006_a}.  Furthermore, a surprisingly universal empirical scaling relation between the Hall and longitudinal conductivities,  $\sigma_{xy} \sim \sigma_{xx}^{\gamma}$, $\gamma \approx 1.6$ has been found to be obeyed by a number of materials on the lower side of their conductivity values \cite{Fukumura:2007_a}, where Anderson-Mott quantum localization effects should be important.

In this Letter, we report on  Hall resistance studies as a function of temperature and gate electric field carried out for metal-insulator-semiconductor (MIS) structures containing a thin conducting channel of ferromagnetic (Ga,Mn)As. We find out that in the  $\sigma_{xx}$ range up to $~10^2$~S/cm, $\sigma_{xy}$ obeys a scaling relation with a similar value of the exponent $\gamma$.  However, for $\sigma_{xx} \gtrsim 10^2$~S/cm the scaling relation breaks down entirely. Surprisingly, in this regime and below the Curie temperature $T_{\mathrm{C}}$, $\sigma_{xy}$  tends to decrease rather abruptly with decreasing temperature, and even reverses its sign in some cases, in the region where neither resistance $R$ nor magnetization $M$ vary significantly with temperature. The effect has not been observed in thicker films and appears to have no explanation within the existing theory, pointing to the importance of yet unrevealed confinement effects.

The studied thin layers of tensile-strained (Ga,Mn)As have been deposited by low-temperature molecular beam epitaxy onto a buffer layer consisting of 4-nm GaAs/ 30-nm Al$_{0.75}$Ga$_{0.25}$As/~500-nm In$_{0.15}$Ga$_{0.85}$As/30-nm GaAs grown on a semi-insulating GaAs (001) substrate.  Upon growth, Hall bars having a channel of 30 or 40-$\mu$m width and $\sim$200-$\mu$m length are patterned by photolithography and wet etching. Subsequently, samples are annealed at 180$^{\circ}$C for 5~min or introduced directly into an atomic layer deposition chamber, where a 40-nm-thick oxide insulator is deposited at a substrate temperature of 120-150$^{\circ}$C.  Finally, 5-nm Cr/ 100-nm Au gate electrode is formed. Owing to the tensile strain, the easy axis is perpendicular to the plane, so that the height of the Hall voltage hystereses provides directly the magnitude of the anomalous Hall resistance $R_{yx}$.

Altogether 18 MIS structures, numbered from 1 to 18, have been investigated. As tabulated in the supplementary material\cite{EPAPS}, they differ in nominal Mn composition ($3\% \leq x \leq 17.5\%$), thickness (3.5~nm $\leq t \leq$ 5~nm), and crystallographic orientation ([110] vs. [$\bar{1}$10]) of the Ga$_{1-x}$Mn$_{x}$As channel, as well as contain three kinds of gate insulators (Al$_2$O$_3$, HfO$_2$, and ZrO$_2$). For this set of samples, and in the employed gate electric field range (5~MV/cm $\geq E_{\mathrm{G}} \geq -5$~MV/cm), $\sigma_{xx}$ spans over 3 orders of magnitude, and the corresponding $T_{\mathrm{C}}$ varies between 35 and 165~K. The lateral homogeneity of the grown wafers is proved by results displayed in Fig. S1 in EPAPS, whereas the data in Fig. S2 demonstrate that the magnitudes of  $T_{\mathrm{C}}$ in the MIS structures and thick layers are virtually identical. Since $t$ is larger than the mean free path but shorter than the phase coherence length of holes in (Ga,Mn)As, whose lower limit is provided by the conductance studies in 1D systems \cite{Wagner:2006_a,Vila:2007_a}, the density of states preserves a 3D shape, whereas localization phenomena acquire a 2D character.

As an example of experimental findings, Figs.~1(a) and 1(b) show the temperature dependence of remanent Hall resistance $R_{yx}$ and sheet resistance $R_{xx}$ at various gate electric fields for the MIS structure containing a 5-nm thick Ga$_{0.949}$Mn$_{0.051}$As channel and the Al$_2$O$_3$ gate insulator. In the inset, the results of  the Hall measurements as a function of the external magnetic field $H$ at 10~K are presented. The squareness of the hysteresis loop indicates the perpendicular-to-plane orientation of the magnetization easy axis, whereas its counter-clockwise chirality, as in the case of an ordinary $M$ vs. $H$ loop, demonstrates that the anomalous Hall coefficient is positive in this case. The vertical axis of Fig.~1(a) presents $R_{yx}(T)$, as determined by hysteresis heights at zero magnetic field. As seen, $T_{\mathrm{C}}$ of the channel layer increases (decreases) by the application of negative (positive) gate electric field which accumulates (depletes) holes in the channel, as witnessed in Fig.~1(b) by a corresponding decrease (increase) of $R_{xx}$. Importantly, according to Fig.~1(a) also the low-temperature values of $R_{yx}$ increase with $R_{xx}$ changed by the gate electric field. Figure 1(c) shows $\sigma_{xy} = R_{yx}/[t(R_{xx}^2 + R_{yx}^2)]$ as a function of $\sigma_{xx} = R_{xx}/[t(R_{xx}^2 + R_{yx}^2)]$ under three different values of $E_{\mathrm{G}}$ at $T = 10$--$30$~K $\lesssim T_{\mathrm{C}}/2$, where the spontaneous spin splitting should be only weakly temperature dependent. The dotted line is a guide for the eyes, which indicates that the values of $\sigma_{xy}$ fall on a single curve, verifying an empirical scaling behavior with $\gamma \approx 1.6$ for this structure. In 12 samples of all samples, $\gamma$ is found to be within the range of $1.6\pm 0.4$, showing virtually no temperature dependence below $T_{\mathrm{C}}/2$ \cite{EPAPS}.

\begin{figure}[b]
\includegraphics[width=3.3in]{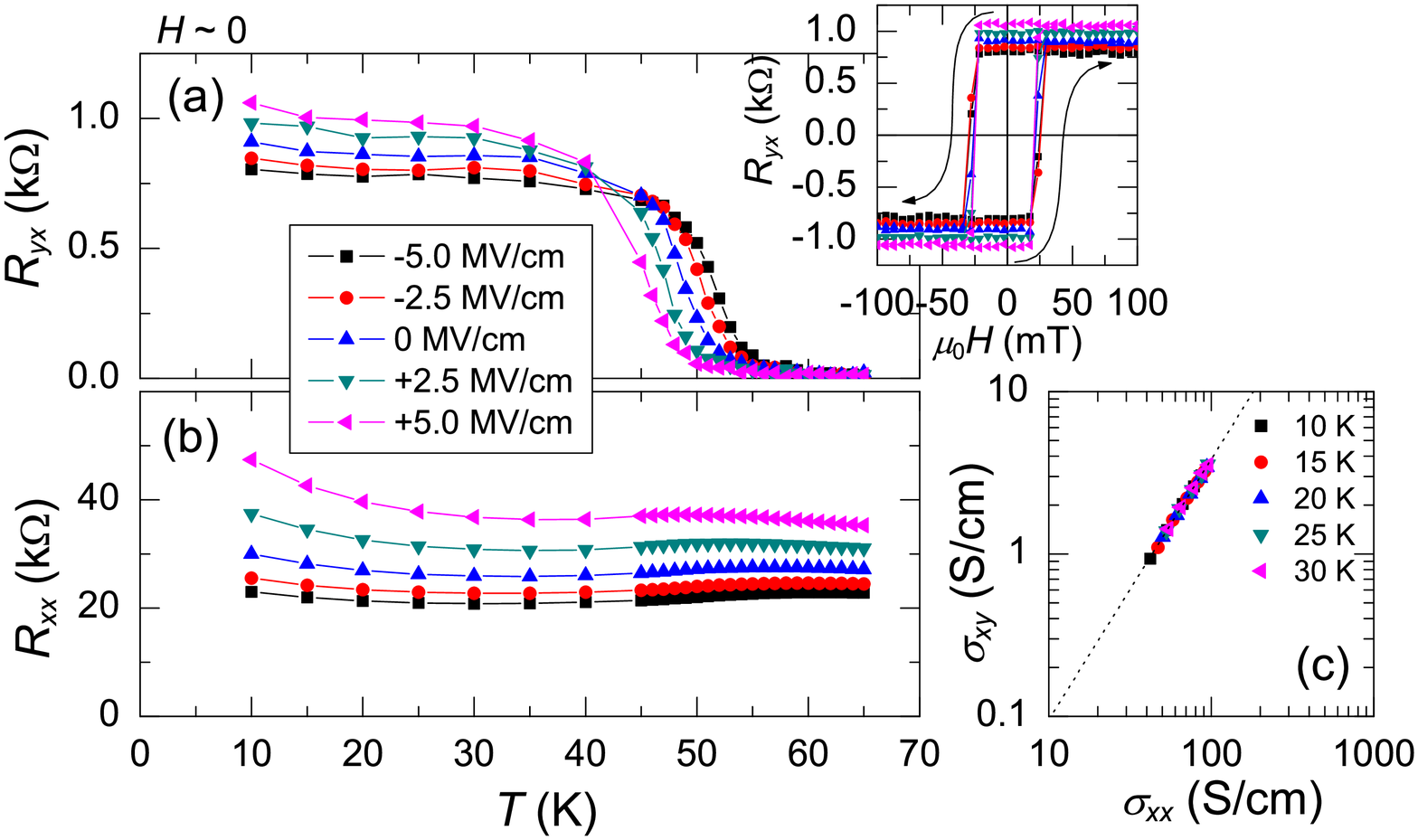}\vspace{-4mm}
\caption{[Color online] (a) Hall and (b) longitudinal sheet resistances of 5-nm thick Ga$_{0.949}$Mn$_{0.051}$As (sample 3) at various gate electric fields. Inset to (a) shows hysteresis loops of the Hall resistance. (c) Relation between Hall and longitudinal conductivities obeying a simple scaling law shown by dotted line.} \label{fig:1}
\end{figure}

Experimental results illustrating an entirely different and so-far not reported behavior of $R_{yx}(T)$ are presented in Fig.~2(a) for the MIS structure containing a 4-nm thick Ga$_{0.875}$Mn$_{0.125}$As channel and the HfO$_2$ gate insulator. While, according to  Fig.~2(b) the temperature dependence of $R_{xx}$ is rather standard,  $R_{yx}$ shows a nonmonotonic temperature dependence attaining a maximum at about 20~K below $T_{\mathrm{C}} \approx 120$~K, followed by a decrease of $R_{yx}$ towards zero with lowering temperature, culminating in a change of sign at $E_{\mathrm{G}} > 1.5$~MV/cm at 10~K, as shown in detail in Fig.~2(c). The data on $R_{yx}(T)$ are to be contrasted with the behavior of magnetization which, according to the results displayed in Fig.~2(d), shows a monotonic increase upon cooling. Interestingly, while the magnitude of $T_{\mathrm{C}}$ in our MIS structures attains similar values as those in thicker films, a non-Brillouin character of $M(T)$ indicates that magnetization fluctuations are rather strongly enhanced by the confinement.

\begin{figure}[b]
\includegraphics[width=3.0in]{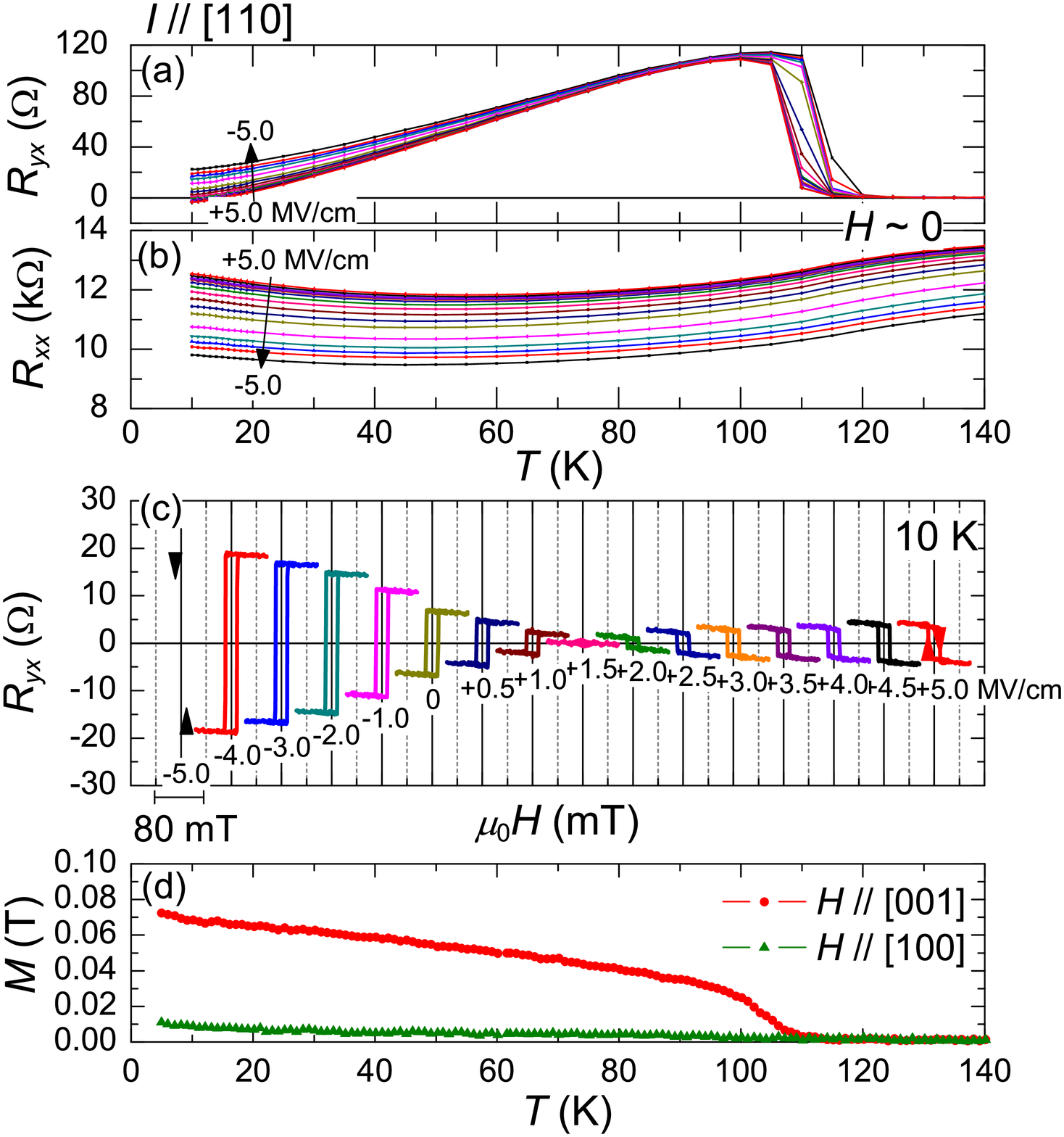}\vspace{-4mm}
\caption{[Color online] (a) Hall and (b) longitudinal sheet resistances of 4-nm thick Ga$_{0.875}$Mn$_{0.125}$As (sample 16) at various gate electric fields. (c) Hysteresis loops at various gate electric fields documenting the sign change of the anomalous Hall effect. (d) Temperature dependence of remanent magnetization without gate bias. } \label{fig:2}
\end{figure}

The findings for this and other samples showing similar properties, collected in Fig.~3, demonstrate clearly that the relation between the Hall resistance and magnetization or carrier polarization can be rather complex. In particular, despite that $R_{xx}$ is virtually temperature independent and the hole liquid degenerate,  $R_{yx}$ and, thus, $\sigma_{xy}$ decrease abruptly or even change sign on lowering temperature, as shown in Figs.~2 and 3. In this range, the temperature derivatives of $|R_{yx}(T)|$ and $M(T)$ acquire opposite signs, indicating that Hall measurements for temperature dependent magnetometry should be applied with care. Furthermore, the revealed change of sign of $R_{yx}$ calls into question the generality of a simple scaling formula.

\begin{figure}[t]
\includegraphics[width=3.3in]{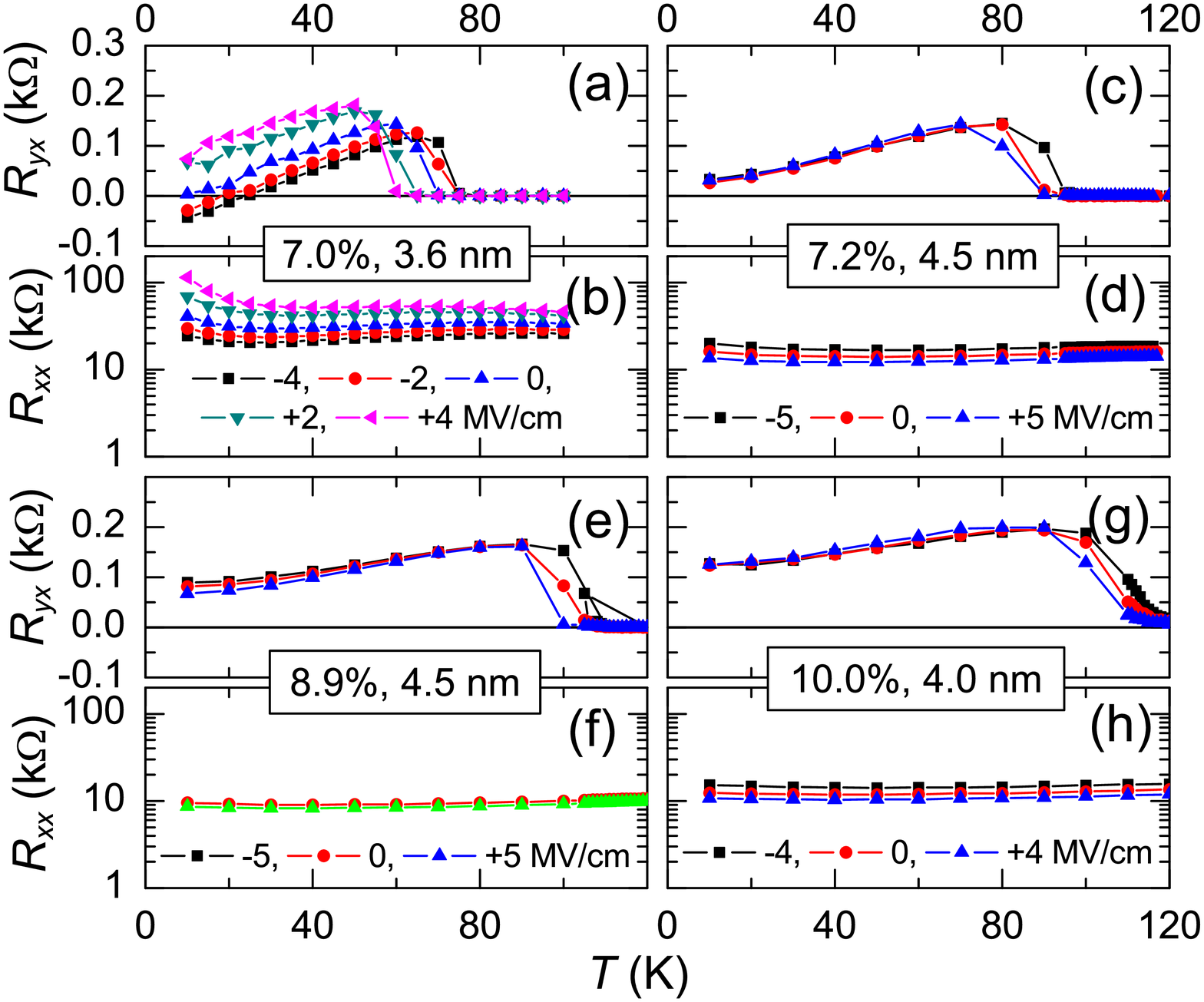}\vspace{-4mm}
\caption{[Color online]  Hall $R_{yx}$ and longitudinal $R_{xx}$ sheet resistances at various values of the gate electric fields in MIS structures [(a) and (b) for sample 11, (c) and (d) for sample 12, (e) and (f) for sample 14, and (g) and (h) for sample 15] showing a strong temperature dependence of $R_{yx}$, similar to the one presented in Fig.~2.} \label{fig:3}
\end{figure}

According to the recent theory of the intrinsic AHE in (Ga,Mn)As \cite{Werpachowska:2009_a}, an appropriately strong tensile strain or the bulk inversion asymmetry--- the Dresselhaus effect\cite{Dresselhaus}---can result in a negative sign of the AHE for a sufficiently small magnitude of scattering broadening $\Gamma$. In fact, a negative sign of the $R_{yx}$ has been observed in films of (In,Mn)As \cite{Munekata:1997_a,Matsukura:2002_c}, (In,Mn)Sb \cite{Yanagi:2004_a,Mihaly:2008_a}, and (Ga,Mn)Sb \cite{Eginligil:2008_a}, where the bulk asymmetry is expected to be much stronger. However, the striking behavior of $R_{yx}(T)$, including the change of sign revealed here, has not been anticipated theoretically [4,5]. It appears to be related to the reduced dimensionality, as it has not been observed in thicker (Ga,Mn)As films grown by us with similar conductivity $T_{\mathrm{C}}$ and strain \cite{Chiba:2006_a,Pu:2008_a}.

In an attempt to interpret our findings we note that thin layers are expected to exhibit structure inversion asymmetry constituting an additional source of spin-orbit coupling. Despite that the gate electric field does not affect significantly the temperature dependence of $R_{yx}$, as seen in Fig.~3, the presence of two different interfaces can account for the structure asymmetry and the corresponding lowering of the point symmetry from  $D_{2d}$ to $C_{2v}$. In order to explain the data within this scenario,  the structure asymmetry  contribution to $R_{yx}$ should be negative and its amplitude take over the bulk terms at sufficiently high magnetization values.

\begin{figure}[t]
\includegraphics[width=3.3in]{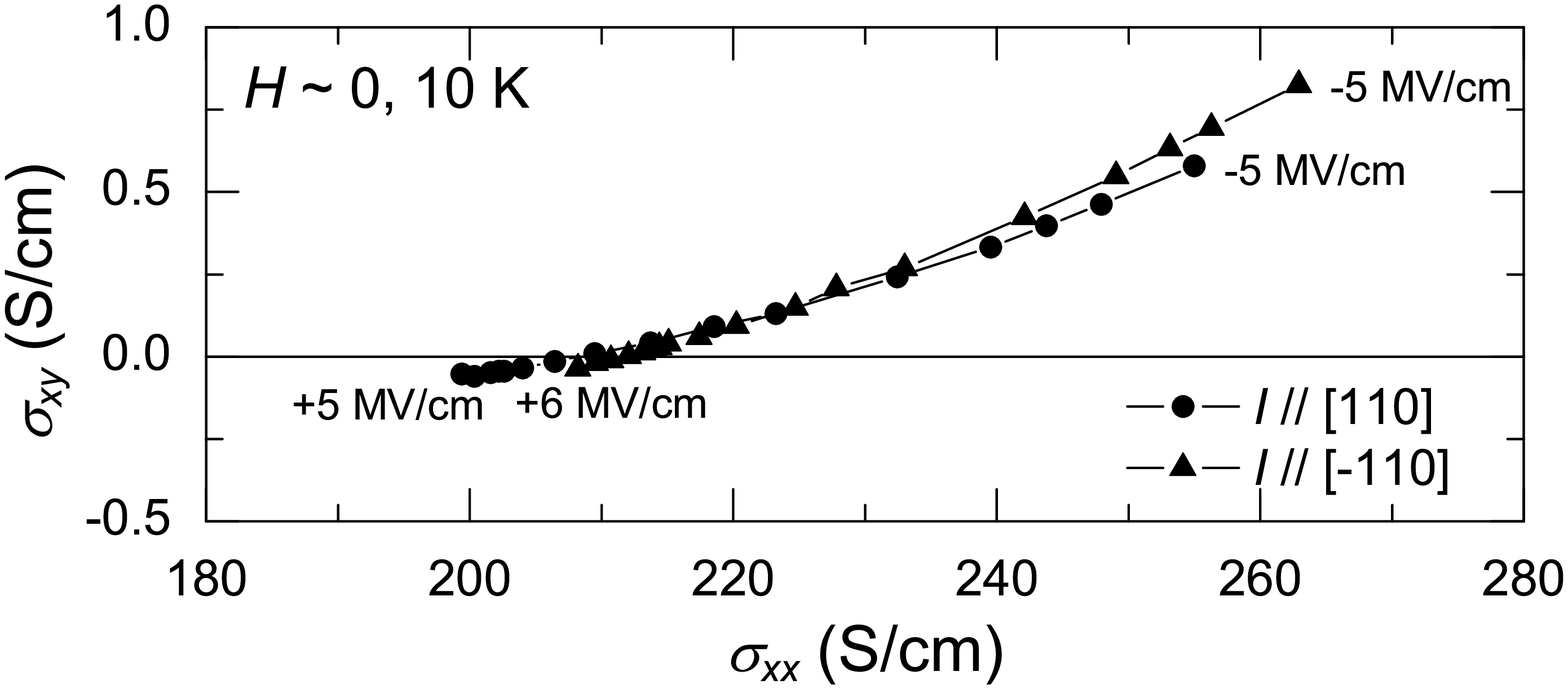}\vspace{-4mm}
\caption{Hall conductivity vs. conductivity for 4-nm thick Ga$_{0.875}$Mn$_{0.125}$As channels oriented along [110] (sample 16) and [$\bar{1}$10] (sample 17) in-plane crystallographic orientations.} \label{fig:4}
\end{figure}

Since the in-plane anisotropy of the conductivity tensor would provide a direct proof that the symmetry is lowered to $C_{2v}$, we have examined MIS structures originating from the same wafer but having the channels oriented along either [110] or [$\bar{1}$10] crystal axis. As shown in Fig.~4, similar values of Hall and longitudinal conductance are observed in both cases. Though the negative result of this experiment does not provide a support for the structural asymmetry model, it does not disprove it, as an accidental degeneracy cannot be excluded ($e.g.$, due to a compensation of asymmetry in the values of the density of states and relaxation time).

Another possibility is that dimensional quantization of the transverse motion leads to a significant reconstruction of the topology of the Fermi surface introducing, in particular, a number of additional subband crossings.  This scenario may explain why the negative sign of $\sigma_{xy}$ appears on depleting the channel in one structure [Fig.~2(a)], whereas it shows up in the accumulation regime in another sample [Fig.~3(a)]. A detailed Fermi sheet behavior in this region ($e.g.$, crossing vs. anticrossing) may be sensitive to symmetry lowering and phase breaking mechanisms such as spin-orbit as well as temperature dependent inelastic and spin-disorder scattering. At the same time, the confinement-induced upward shift of the hole energies may enhance the importance of the Dresselhaus contribution determined by the admixture of the conduction band wave functions. Within this scenario, the behavior of $\sigma_{xy}$ as a function of temperature reflects a subtle and spin polarization-dependent balance between positive and negative terms originating from the intra-atom and bulk inversion asymmetry electric fields, respectively.

In Fig.~5 we summarize our findings by reporting $\sigma_{xy}$ as a function of $\sigma_{xx}$ at 10 K for various gate electric fields. The results \cite{Matsukura:1998_a,Chiba:2006_a,Pu:2008_a} obtained previously for thick films are also shown for comparison. Despite the fact that the particular samples may differ in magnetization values---which may likely affect the form of the scaling \cite{Chun:2007_a}---we see that over a wide range of conductivities up to $10^2$~S/cm the relation  $\sigma_{xy} \sim \sigma_{xx}^{\gamma}$ is obeyed and implies  $\gamma\sim 1.6$, in agreement with the value $\gamma = 1.5$ found for thick films of (Ga,Mn)As in the magnetic field \cite{Shen:2008_a}. However, the scaling relation breaks down rather severely in the case of MIS structures, as discussed above. We also note that $\gamma$ determined at each temperature has almost no temperature dependence below $T_\mathrm{C}/2$, according to the results shown in Fig. S2 of the supplementaly material. However, on approaching $T_\mathrm{C}$ we have found the scaling to be not obeyed (Fig. S4).

\begin{figure}[t]
\includegraphics[width=3.3in]{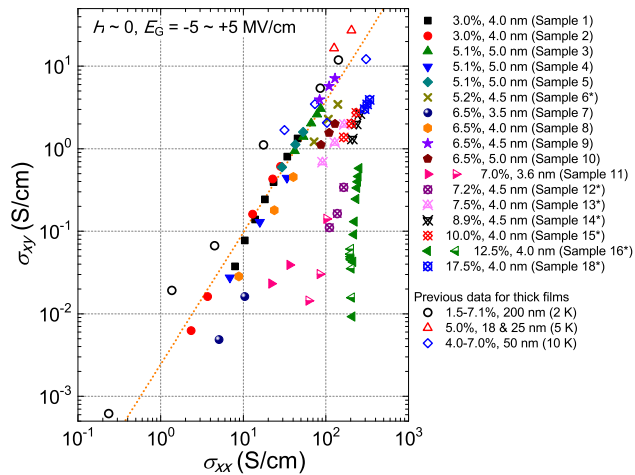}\vspace{-4mm}
\caption{[Color online] Relation between Hall and longitudinal conductivities at various values of gate electric fields $ E_{\mathrm{G}}$ at 10 K (30 K for Sample 1) for the studied MIS structures of Ga$_{1-x}$Mn$_{x}$As. The legend shows the values of the Mn content $x$, channel thickness $t$, as well as the sample number. The asterisks (*) mark the samples which were annealed before deposition of insulator layer. Half-filled symbols show the absolute value of negative $\sigma_{xy}$. The previous data for thick (Ga,Mn)As films are shown by open symbols (circles \cite{Matsukura:1998_a}, triangles up \cite{Chiba:2006_a}, and diamonds \cite{Pu:2008_a}). The dotted line shows the dependence $\sigma_{xy} \sim \sigma_{xx}^{\gamma}$, $\gamma = 1.6$.} \label{fig:5}
\end{figure}

We note that the empirical scaling found for thick films as well as for thin layers with low conductance cannot be explained by the intrinsic mechanisms of the AHE, as the corresponding theory predicts a decrease of $\sigma_{xy}$ with both hole density \cite{Jungwirth:2002_a,Werpachowska:2009_a,Dietl:2003_c} and scattering time \cite{Werpachowska:2009_a,Jungwirth:2003_b} in a wide range of relevant hole concentrations and spin-splittings. This suggests that the physics of AHE is dominated by the proximity to the Anderson-Mott localization boundary \cite{Belitz:1994_a,Dietl:2008_c}. So far the influence of quantum interference effects on the anomalous Hall conductance have been studied considering the side-jump and skew-scattering terms within the single-particle theory \cite{Dugaev:2001_a}. The data summarized in Fig.~5 call for the extension of the theory towards the intrinsic AHE with the effects of disorder on interference of carrier-carrier scattering amplitudes taken into account.

In summary, we have found that the anomalous Hall effect of the magnetic semiconductor (Ga,Mn)As acquires qualitatively new and not anticipated features when dimensionality and disorder are reduced. The revealed striking temperature dependence of the Hall conductance indicates a significant contribution of confinement phenomena. At higher levels of disorder, the scaling relation between $\sigma_{xy}$ and $\sigma_{xx}$, similar to that observed previously for thicker samples and so-far unexplained, is recovered. 

The authors thank S. Murakami and N. Nagaosa for useful discussions, and L. Ye for experimental support. The work in Sendai was supported in part by Grant-in-Aids from MEXT/JSPS, the GCOE Program at Tohoku University, and the Research and Development for Next-Generation Information Technology Program from MEXT, whereas the work in Warsaw was supported by the President of the Polish Academy of Sciences for doctoral students (A. W.) and by FunDMS Advanced Grant of European Research Council within "Ideas" 7th Framework Programme of EC (T. D.).\newline


\vspace{-9mm}
\begin{thebibliography}{26}
\expandafter\ifx\csname natexlab\endcsname\relax\def\natexlab#1{#1}\fi
\expandafter\ifx\csname bibnamefont\endcsname\relax
  \def\bibnamefont#1{#1}\fi
\expandafter\ifx\csname bibfnamefont\endcsname\relax
  \def\bibfnamefont#1{#1}\fi
\expandafter\ifx\csname citenamefont\endcsname\relax
  \def\citenamefont#1{#1}\fi
\expandafter\ifx\csname url\endcsname\relax
  \def\url#1{\texttt{#1}}\fi
\expandafter\ifx\csname urlprefix\endcsname\relax\def\urlprefix{URL }\fi
\providecommand{\bibinfo}[2]{#2}
\providecommand{\eprint}[2][]{\url{#2}}

\bibitem[{\citenamefont{Sundaram and Niu}(1999)}]{Sundaram:1999_a}
\bibinfo{author}{\bibfnamefont{G.}~\bibnamefont{Sundaram}} \bibnamefont{and}
  \bibinfo{author}{\bibfnamefont{Q.}~\bibnamefont{Niu}},
  \bibinfo{journal}{Phys. Rev. B} \textbf{\bibinfo{volume}{59}},
  \bibinfo{pages}{14915} (\bibinfo{year}{1999}).

\bibitem[{\citenamefont{Jungwirth et~al.}(2002)\citenamefont{Jungwirth, Niu,
  and MacDonald}}]{Jungwirth:2002_a}
\bibinfo{author}{\bibfnamefont{T.}~\bibnamefont{Jungwirth}},
  \bibinfo{author}{\bibfnamefont{Q.}~\bibnamefont{Niu}}, \bibnamefont{and}
  \bibinfo{author}{\bibfnamefont{A.}~\bibfnamefont{H.}~\bibnamefont{MacDonald}},
  \bibinfo{journal}{Phys. Rev. Lett.} \textbf{\bibinfo{volume}{88}},
  \bibinfo{pages}{207208} (\bibinfo{year}{2002}).


\bibitem[{\citenamefont{Jungwirth et~al.}(2006)\citenamefont{Jungwirth, Sinova,
  {Ma\v{s}ek}, {Ku\v{c}era}, and MacDonald}}]{Jungwirth:2006_a}
\bibinfo{author}{\bibfnamefont{T.}~\bibnamefont{Jungwirth}},
  \bibinfo{author}{\bibfnamefont{J.}~\bibnamefont{Sinova}},
  \bibinfo{author}{\bibfnamefont{J.}~\bibnamefont{{Ma\v{s}ek}}},
  \bibinfo{author}{\bibfnamefont{J.}~\bibnamefont{{Ku\v{c}era}}},
  \bibnamefont{and} \bibinfo{author}{\bibfnamefont{A.~H.}
  \bibnamefont{MacDonald}}, \bibinfo{journal}{Rev. Mod. Phys.}
  \textbf{\bibinfo{volume}{78}}, \bibinfo{pages}{809} (\bibinfo{year}{2006}).

  \bibitem{Nagaosa:2009_a} N. Nagaosa, J. Sinova, S. Onoda, A. H. MacDonald, and N. P. Ong, arXiv:0904.4154.


\bibitem[{\citenamefont{Werpachowska and Dietl}(2009)}]{Werpachowska:2009_a}
\bibinfo{author}{\bibfnamefont{A.}~\bibnamefont{Werpachowska}}
  \bibnamefont{and} \bibinfo{author}{\bibfnamefont{T.}~\bibnamefont{Dietl}},
  \bibinfo{journal}{arXiv:0910.1907}  (\bibinfo{year}{2009}).

\bibitem[{\citenamefont{Rashba}(2004)}]{Rashba:2004_d}
\bibinfo{author}{\bibfnamefont{E.~I.} \bibnamefont{Rashba}},
  \bibinfo{journal}{Phys. Rev. B} \textbf{\bibinfo{volume}{70}},
  \bibinfo{pages}{201309(R)} (\bibinfo{year}{2004}).

\bibitem[{\citenamefont{Inoue et~al.}(2006)\citenamefont{Inoue, Kato, Ishikawa,
  Itoh, Bauer, and Molenkamp}}]{Inoue:2006_a}
\bibinfo{author}{\bibfnamefont{J.}~\bibfnamefont{I.}~\bibnamefont{Inoue}},
  \bibinfo{author}{\bibfnamefont{T.}~\bibnamefont{Kato}},
  \bibinfo{author}{\bibfnamefont{Y.}~\bibnamefont{Ishikawa}},
  \bibinfo{author}{\bibfnamefont{H.}~\bibnamefont{Itoh}},
  \bibinfo{author}{\bibfnamefont{G.~E.~W.} \bibnamefont{Bauer}},
  \bibnamefont{and} \bibinfo{author}{\bibfnamefont{L.~W.}
  \bibnamefont{Molenkamp}}, \bibinfo{journal}{Phys. Rev. Lett.}
  \textbf{\bibinfo{volume}{97}}, \bibinfo{pages}{046604}
  (\bibinfo{year}{2006}).

\bibitem[{\citenamefont{Fukumura et~al.}(2007)\citenamefont{Fukumura, Toyosaki,
  Ueno, Nakano, Yamasaki, and Kawasaki}}]{Fukumura:2007_a}
\bibinfo{author}{\bibfnamefont{T.}~\bibnamefont{Fukumura}},
  \bibinfo{author}{\bibfnamefont{H.}~\bibnamefont{Toyosaki}},
  \bibinfo{author}{\bibfnamefont{K.}~\bibnamefont{Ueno}},
  \bibinfo{author}{\bibfnamefont{M.}~\bibnamefont{Nakano}},
  \bibinfo{author}{\bibfnamefont{T.}~\bibnamefont{Yamasaki}}, \bibnamefont{and}
  \bibinfo{author}{\bibfnamefont{M.}~\bibnamefont{Kawasaki}},
  \bibinfo{journal}{Jpn. J. Appl. Phys.} \textbf{\bibinfo{volume}{46}},
  \bibinfo{pages}{L642} (\bibinfo{year}{2007}).

\bibitem{EPAPS} See EPAPS Document No. E-PRLTAO-xxx for a table and figures summarizing sample properties and uniformity. For more information on EPAPS, see http://www.aip.org/pubservs/epaps.html.

\bibitem[{\citenamefont{Wagner et~al.}(2006)\citenamefont{Wagner, Neumaier,
  Reinwald, Wegscheider, and Weiss}}]{Wagner:2006_a}
\bibinfo{author}{\bibfnamefont{K.}~\bibnamefont{Wagner}},
  \bibinfo{author}{\bibfnamefont{D.}~\bibnamefont{Neumaier}},
  \bibinfo{author}{\bibfnamefont{M.}~\bibnamefont{Reinwald}},
  \bibinfo{author}{\bibfnamefont{W.}~\bibnamefont{Wegscheider}},
  \bibnamefont{and} \bibinfo{author}{\bibfnamefont{D.}~\bibnamefont{Weiss}},
  \bibinfo{journal}{Phys. Rev. Lett.} \textbf{\bibinfo{volume}{97}},
  \bibinfo{pages}{056803} (\bibinfo{year}{2006}).
  
 \bibitem{Vila:2007_a} 
 L. Vila, R. Giraud, L. Thevenard, A. Lemaitre, F. Pierre, J. Dufouleur, D. Mailly, B.
Barbara, and G. Faini, Phys. Rev. Lett. {\bf 98}, 027204 (2007).

\bibitem[{\citenamefont{Munekata et~al.}(1997)\citenamefont{Munekata, Abe,
  Koshihara, Oiwa, Hirasawa, Katsumoto, Iye, Urano, and
  Takagi}}]{Munekata:1997_a}
\bibinfo{author}{\bibfnamefont{H.}~\bibnamefont{Munekata}},
  \bibinfo{author}{\bibfnamefont{T.}~\bibnamefont{Abe}},
  \bibinfo{author}{\bibfnamefont{S.}~\bibnamefont{Koshihara}},
  \bibinfo{author}{\bibfnamefont{A.}~\bibnamefont{Oiwa}},
  \bibinfo{author}{\bibfnamefont{M.}~\bibnamefont{Hirasawa}},
  \bibinfo{author}{\bibfnamefont{S.}~\bibnamefont{Katsumoto}},
  \bibinfo{author}{\bibfnamefont{Y.}~\bibnamefont{Iye}},
  \bibinfo{author}{\bibfnamefont{C.}~\bibnamefont{Urano}}, \bibnamefont{and}
  \bibinfo{author}{\bibfnamefont{H.}~\bibnamefont{Takagi}},
  \bibinfo{journal}{J. Appl. Phys.} \textbf{\bibinfo{volume}{81}},
  \bibinfo{pages}{4862} (\bibinfo{year}{1997}).

\bibitem[{\citenamefont{Matsukura et~al.}(2002)\citenamefont{Matsukura, Chiba,
  Omiya, Abe, Dietl, Ohno, Ohtani, and Ohno}}]{Matsukura:2002_c}
\bibinfo{author}{\bibfnamefont{F.}~\bibnamefont{Matsukura}},
  \bibinfo{author}{\bibfnamefont{D.}~\bibnamefont{Chiba}},
  \bibinfo{author}{\bibfnamefont{T.}~\bibnamefont{Omiya}},
  \bibinfo{author}{\bibfnamefont{E.}~\bibnamefont{Abe}},
  \bibinfo{author}{\bibfnamefont{T.}~\bibnamefont{Dietl}},
  \bibinfo{author}{\bibfnamefont{Y.}~\bibnamefont{Ohno}},
  \bibinfo{author}{\bibfnamefont{K.}~\bibnamefont{Ohtani}}, \bibnamefont{and}
  \bibinfo{author}{\bibfnamefont{H.}~\bibnamefont{Ohno}},
  \bibinfo{journal}{Physica E} \textbf{\bibinfo{volume}{12}},
  \bibinfo{pages}{351} (\bibinfo{year}{2002}).

\bibitem[{\citenamefont{Dresselhaus}(1955)}]{Dresselhaus}
\bibinfo{author}{\bibfnamefont{G.}~\bibnamefont{Dresselhaus}}, \bibinfo{journal}{Phys. Rev.} \textbf{\bibinfo{volume}{100}}, \bibinfo{pages}{580} 
  (\bibinfo{year}{1955}).  

\bibitem[{\citenamefont{Yanagi et~al.}(2004)\citenamefont{Yanagi, Kuga,
  Slupinski, and Munekata}}]{Yanagi:2004_a}
\bibinfo{author}{\bibfnamefont{S.}~\bibnamefont{Yanagi}},
  \bibinfo{author}{\bibfnamefont{K.}~\bibnamefont{Kuga}},
  \bibinfo{author}{\bibfnamefont{T.}~\bibnamefont{Slupinski}},
  \bibnamefont{and} \bibinfo{author}{\bibfnamefont{H.}~\bibnamefont{Munekata}},
  \bibinfo{journal}{Physica E} \textbf{\bibinfo{volume}{20}},
  \bibinfo{pages}{333} (\bibinfo{year}{2004}).

\bibitem[{\citenamefont{Mih\'aly et~al.}(2008)\citenamefont{Mih\'aly, Csontos,
  Bord\'acs, K\'ezsm\'arki, Wojtowicz, Liu, Jank\'o, and
  Furdyna}}]{Mihaly:2008_a}
\bibinfo{author}{\bibfnamefont{G.}~\bibnamefont{Mih\'aly}},
  \bibinfo{author}{\bibfnamefont{M.}~\bibnamefont{Csontos}},
  \bibinfo{author}{\bibfnamefont{S.}~\bibnamefont{Bord\'acs}},
  \bibinfo{author}{\bibfnamefont{I.}~\bibnamefont{K\'ezsm\'arki}},
  \bibinfo{author}{\bibfnamefont{T.}~\bibnamefont{Wojtowicz}},
  \bibinfo{author}{\bibfnamefont{X.}~\bibnamefont{Liu}},
  \bibinfo{author}{\bibfnamefont{B.}~\bibnamefont{Jank\'o}}, \bibnamefont{and}
  \bibinfo{author}{\bibfnamefont{J.~K.} \bibnamefont{Furdyna}},
  \bibinfo{journal}{Phys. Rev. Lett.} \textbf{\bibinfo{volume}{100}},
  \bibinfo{pages}{107201} (\bibinfo{year}{2008}).

\bibitem[{\citenamefont{Eginligil et~al.}(2008)\citenamefont{Eginligil, Kim,
  Yoon, Bird, Luo, and McCombe}}]{Eginligil:2008_a}
\bibinfo{author}{\bibfnamefont{M.}~\bibnamefont{Eginligil}},
  \bibinfo{author}{\bibfnamefont{G.}~\bibnamefont{Kim}},
  \bibinfo{author}{\bibfnamefont{Y.}~\bibnamefont{Yoon}},
  \bibinfo{author}{\bibfnamefont{J.~P.} \bibnamefont{Bird}},
  \bibinfo{author}{\bibfnamefont{H.}~\bibnamefont{Luo}}, \bibnamefont{and}
  \bibinfo{author}{\bibfnamefont{B.~D.} \bibnamefont{McCombe}},
  \bibinfo{journal}{Physica E} \textbf{\bibinfo{volume}{40}},
  \bibinfo{pages}{2104} (\bibinfo{year}{2008}).

\bibitem[{\citenamefont{Chiba et~al.}(2006)\citenamefont{Chiba, Yamanouchi,
  Matsukura, Dietl, and Ohno}}]{Chiba:2006_a}
\bibinfo{author}{\bibfnamefont{D.}~\bibnamefont{Chiba}},
  \bibinfo{author}{\bibfnamefont{M.}~\bibnamefont{Yamanouchi}},
  \bibinfo{author}{\bibfnamefont{F.}~\bibnamefont{Matsukura}},
  \bibinfo{author}{\bibfnamefont{T.}~\bibnamefont{Dietl}}, \bibnamefont{and}
  \bibinfo{author}{\bibfnamefont{H.}~\bibnamefont{Ohno}},
  \bibinfo{journal}{Phys. Rev. Lett.} \textbf{\bibinfo{volume}{96}},
  \bibinfo{pages}{096602} (\bibinfo{year}{2006}).

\bibitem[{\citenamefont{Pu et~al.}(2008)\citenamefont{Pu, Chiba, Matsukura,
  Ohno, and Shi}}]{Pu:2008_a}
\bibinfo{author}{\bibfnamefont{Y.}~\bibnamefont{Pu}},
  \bibinfo{author}{\bibfnamefont{D.}~\bibnamefont{Chiba}},
  \bibinfo{author}{\bibfnamefont{F.}~\bibnamefont{Matsukura}},
  \bibinfo{author}{\bibfnamefont{H.}~\bibnamefont{Ohno}}, \bibnamefont{and}
  \bibinfo{author}{\bibfnamefont{J.}~\bibnamefont{Shi}},
  \bibinfo{journal}{Phys. Rev. Lett.} \textbf{\bibinfo{volume}{101}},
  \bibinfo{pages}{117208} (\bibinfo{year}{2008}).

\bibitem[{\citenamefont{Matsukura et~al.}(1998)\citenamefont{Matsukura, Ohno,
  Shen, and Sugawara}}]{Matsukura:1998_a}
\bibinfo{author}{\bibfnamefont{F.}~\bibnamefont{Matsukura}},
  \bibinfo{author}{\bibfnamefont{H.}~\bibnamefont{Ohno}},
  \bibinfo{author}{\bibfnamefont{A.}~\bibnamefont{Shen}}, \bibnamefont{and}
  \bibinfo{author}{\bibfnamefont{Y.}~\bibnamefont{Sugawara}},
  \bibinfo{journal}{Phys. Rev. B} \textbf{\bibinfo{volume}{57}},
  \bibinfo{pages}{R2037} (\bibinfo{year}{1998}).

\bibitem[{\citenamefont{Chun et~al.}(2007)\citenamefont{Chun, Kim, Choi, Jeong,
  Lee, Suh, Oh, Kim, Khim, Woo et~al.}}]{Chun:2007_a}
\bibinfo{author}{\bibfnamefont{S.~H.} \bibnamefont{Chun}},
  \bibinfo{author}{\bibfnamefont{Y.~S.} \bibnamefont{Kim}},
  \bibinfo{author}{\bibfnamefont{H.~K.} \bibnamefont{Choi}},
  \bibinfo{author}{\bibfnamefont{I.~T.} \bibnamefont{Jeong}},
  \bibinfo{author}{\bibfnamefont{W.~O.} \bibnamefont{Lee}},
  \bibinfo{author}{\bibfnamefont{K.~S.} \bibnamefont{Suh}},
  \bibinfo{author}{\bibfnamefont{Y.~S.} \bibnamefont{Oh}},
  \bibinfo{author}{\bibfnamefont{K.~H.} \bibnamefont{Kim}},
  \bibinfo{author}{\bibfnamefont{Z.~G.} \bibnamefont{Khim}},
  \bibinfo{author}{\bibfnamefont{J.~C.} \bibnamefont{Woo}},
  \bibnamefont{et~al.}, \bibinfo{journal}{Phys. Rev. Lett.}
  \textbf{\bibinfo{volume}{98}}, \bibinfo{pages}{026601}
  (\bibinfo{year}{2007}).

\bibitem[{\citenamefont{Shen et~al.}(2008)\citenamefont{Shen, Liu, Ge, Furdyna,
  Dobrowolska, and Jaroszynski}}]{Shen:2008_a}
\bibinfo{author}{\bibfnamefont{S.}~\bibnamefont{Shen}},
  \bibinfo{author}{\bibfnamefont{X.}~\bibnamefont{Liu}},
  \bibinfo{author}{\bibfnamefont{Z.}~\bibnamefont{Ge}},
  \bibinfo{author}{\bibfnamefont{J.~K.} \bibnamefont{Furdyna}},
  \bibinfo{author}{\bibfnamefont{M.}~\bibnamefont{Dobrowolska}},
  \bibnamefont{and}
  \bibinfo{author}{\bibfnamefont{J.}~\bibnamefont{Jaroszynski}},
  \bibinfo{journal}{J. Appl. Phys.} \textbf{\bibinfo{volume}{103}},
  \bibinfo{pages}{07D134} (\bibinfo{year}{2008}).

\bibitem[{\citenamefont{Dietl et~al.}(2003)\citenamefont{Dietl, Matsukura,
  Ohno, Cibert, and Ferrand}}]{Dietl:2003_c}
\bibinfo{author}{\bibfnamefont{T.}~\bibnamefont{Dietl}},
  \bibinfo{author}{\bibfnamefont{F.}~\bibnamefont{Matsukura}},
  \bibinfo{author}{\bibfnamefont{H.}~\bibnamefont{Ohno}},
  \bibinfo{author}{\bibfnamefont{J.}~\bibnamefont{Cibert}}, \bibnamefont{and}
  \bibinfo{author}{\bibfnamefont{D.}~\bibnamefont{Ferrand}}, in
  \emph{\bibinfo{booktitle}{Recent Trends in Theory of Physical Phenomena in
  High Magnetic Fields}}, edited by
  \bibinfo{editor}{\bibfnamefont{I.}~\bibnamefont{Vagner}}
  (\bibinfo{publisher}{Kluwer, Dordrecht}, \bibinfo{year}{2003}), p.
  \bibinfo{pages}{197}, \eprint{cond-mat/0306484}.

\bibitem[{\citenamefont{Jungwirth et~al.}(2003)\citenamefont{Jungwirth, Sinova,
  Wang, Edmonds, Campion, Gallagher, Foxon, Niu, and
  MacDonald}}]{Jungwirth:2003_b}
\bibinfo{author}{\bibfnamefont{T.}~\bibnamefont{Jungwirth}},
  \bibinfo{author}{\bibfnamefont{J.}~\bibnamefont{Sinova}},
  \bibinfo{author}{\bibfnamefont{K.~Y.} \bibnamefont{Wang}},
  \bibinfo{author}{\bibfnamefont{K.~W.} \bibnamefont{Edmonds}},
  \bibinfo{author}{\bibfnamefont{R.~P.} \bibnamefont{Campion}},
  \bibinfo{author}{\bibfnamefont{B.~L.} \bibnamefont{Gallagher}},
  \bibinfo{author}{\bibfnamefont{C.~T.} \bibnamefont{Foxon}},
  \bibinfo{author}{\bibfnamefont{Q.}~\bibnamefont{Niu}}, \bibnamefont{and}
  \bibinfo{author}{\bibfnamefont{A.~H.} \bibnamefont{MacDonald}},
  \bibinfo{journal}{Appl. Phys. Lett.} \textbf{\bibinfo{volume}{83}},
  \bibinfo{pages}{320} (\bibinfo{year}{2003}).

\bibitem[{\citenamefont{Belitz and Kirkpatrick}(1994)}]{Belitz:1994_a}
\bibinfo{author}{\bibfnamefont{D.}~\bibnamefont{Belitz}} \bibnamefont{and}
  \bibinfo{author}{\bibfnamefont{T.~R.} \bibnamefont{Kirkpatrick}},
  \bibinfo{journal}{Rev. Mod. Phys.} \textbf{\bibinfo{volume}{66}},
  \bibinfo{pages}{261} (\bibinfo{year}{1994}).

\bibitem[{\citenamefont{Dietl}(2008)}]{Dietl:2008_c}
\bibinfo{author}{\bibfnamefont{T.}~\bibnamefont{Dietl}}, \bibinfo{journal}{J.
  Phys. Soc. Jpn.} \textbf{\bibinfo{volume}{77}}, \bibinfo{pages}{031005}
  (\bibinfo{year}{2008}).

\bibitem[{\citenamefont{Dugaev et~al.}(2001)\citenamefont{Dugaev,
  {Cr\'{e}pieux}, and Bruno}}]{Dugaev:2001_a}
\bibinfo{author}{\bibfnamefont{V.~K.} \bibnamefont{Dugaev}},
  \bibinfo{author}{\bibfnamefont{A.}~\bibnamefont{{Cr\'{e}pieux}}},
  \bibnamefont{and} \bibinfo{author}{\bibfnamefont{P.}~\bibnamefont{Bruno}},
  \bibinfo{journal}{Phys. Rev.} \textbf{\bibinfo{volume}{B 64}},
  \bibinfo{pages}{104411} (\bibinfo{year}{2001}).


\end{thebibliography}

\end{document}